\documentclass[a4paper,fleqn]{cas-dc}
\usepackage[numbers,sort&compress]{natbib}
\def\tsc#1{\csdef{#1}{\textsc{\lowercase{#1}}\xspace}}
\tsc{WGM}
\tsc{QE}
\tsc{EP}
\tsc{PMS}
\tsc{BEC}
\tsc{DE}

\begin{document}
\let\WriteBookmarks\relax
\def\floatpagepagefraction{1}
\def\textpagefraction{.001}
\shorttitle{Characterize noise correlation and enhance coherence via qubit motion}
\shortauthors{Jiaxiu Han et~al.}

\title [mode = title]{Characterize noise correlation and enhance coherence via qubit motion}

\author{Jiaxiu Han}
\credit{Performed the experiment, analyzed
	the data and wrote the manuscript}
\address{Beijing Academy of Quantum Information Sciences, Beijing 100193, China}
\fnmark[1]

\author{Zhiyuan Li}
\credit{Discussed the experiment, and developed the framework of measurement program}
\fnmark[1]

\author{Jingning Zhang}
\credit{Provided theoretical advice}

\author{Huikai Xu}
\credit{Discussed the experiment}

\author{Kehuan Linghu}
\credit{Discussed the experiment}

\author{Yongchao Li}
\credit{Fabricated the devices}

\author{Chengyao Li}
\credit{Fabricated the devices}

\author{Mo Chen}
\credit{Fabricated the devices}

\author{Zhen Yang}
\credit{Fabricated the devices}

\author{Junhua Wang}
\credit{Optimized microwave measurement environment}

\author{Teng Ma}
\credit{Provided theoretical advice}

\author{Guangming Xue}
\credit{Fabricated the devices}
\cormark[1]
\ead{xuegm@baqis.ac.cn}

\author{Yirong Jin}
\cormark[1]
\credit{Discussed the experiment and maintained the dilution refrigerator}
\ead{jinyr@baqis.ac.cn}

\author{Haifeng Yu}
\credit{Supervised the experiment and wrote the manuscript}

\cortext[cor1]{Corresponding author}
\fntext[fn1]{Jiaxiu Han and Zhiyuan Li contributed equally to this work.}

\begin{abstract}
The identification of spacial noise correlation is of critical importance in developing error-corrected quantum devices, but it has barely been studied so far. In this work, we utilize an effective new method called qubit motion, to efficiently determine the noise correlations between any pair of qubits in a 7-qubit superconducting quantum system. The noise correlations between the same pairs of qubits are also investigated when the qubits are at distinct operating frequencies. What's more, in this multi-qubit system with the presence of noise correlations, we demonstrate the enhancing effect of qubit motion on the coherence of logic qubit, and we propose a Motion-CPMG operation sequence to more efficiently protect the logic state from decoherence, which is experimentally demonstrated to extend the decoherence time of logic qubit by nearly one order of magnitude.
\end{abstract}


\maketitle

\section{Introduction}
As one of the most promising platforms to realize quantum computer, superconducting quantum computation (SQC) has reached a level of medium scale with the rapid development in the past two decades. We'll move into the era of realizing the Noisy Intermediate-Scale Quantum (NISQ) device \cite{Preskill_2018} and building logic qubits that allow fault-tolerant quantum computation \cite{Terhal2015Quantum}. Nowadays, developing viable methods to characterize spacial noise correlation has become an important issue, because knowing the correlation is vital for removing unwanted correlated errors \cite{PhysRevLett.109.240504} and performing optimal quantum error correction (QEC) \cite{PhysRevLett.96.050504, chubb2018statistical}. The threshold of error rate for fault-tolerant quantum computation was theoretically predicted to be $10^{-5}$ \cite{aliferis2005quantum} if the environmental noise on individual qubits is uncorrelated. On the contrary, the error rate threshold would drop as low as $10^{-10}$ \cite{PhysRevLett.96.050504} in the presence of spacially correlated noise, because most of the existing QEC codes \cite{548464,gottesman1997stabilizer,bravyi1998quantum,KITAEV20032,PhysRevA.80.052312} are based on the assumption that there is no correlation between qubits, and offer poor protection against correlated errors. One should first characterize the noise correlation in the multi-qubit system, diagnose and understand its source, so as to eliminate or avoid its adverse effect.

Besides, noise results in quantum decoherence. The noise correlations between different qubits in the same quantum chip provide a clue to qualitatively determine the characteristics and source of the noise, so that one can suppress the influence of the noise by different kinds of methods and techniques such as improving circuit layout, and dynamical decoupling \cite{bylander_noise_2011,cai_robust_2012,wang_coherence_2020}, decoherence-free subspaces \cite{Yi_Min_2009,PhysRevA.63.022306,friesen_decoherence-free_2017,altepeter_experimental_2004}, etc, thereby increasing the quantum decoherence time.

In the previous studies, there have been several experimental protocols for determining nearest-neighbor-qubit noise correlation \cite{averin_suppression_2016,PhysRevB.96.115408,PhysRevB.101.235133,brox_bloch-sphere_2012,villar_decoherence_2015,PhysRevApplied.9.064022,PhysRevLett.123.190502,PhysRevB.84.014525,lpke2019twoqubit,xu_experimental_2013} in different quantum systems, but there is few work for characterizing noise correlation between qubits with long-range spacial distances (non-nearest neighbors). A recent work \cite{harper_efficient_2020} proposed a method to construct a quantum noise correlation matrix allowing visualization of correlations between all pairs of qubits, but the method is relatively complicated to implement experimentally. In this article, we achieve the calibration of noise correlation between any two qubits in a one-dimensional qubit chain by an efficient and scalable method called qubit motion \cite{PhysRevLett.116.010501}. The method requires only several single-qubit gates and SWAP gates for the experiment and simple fitting for the construction of quantum noise correlation matrix, which is easy to implement and can be used as a routine to calibrate the noise correlation for the system at any moment. In addition to calibrating the noise correlation, the method of qubit motion has also been demonstrated to increase the decoherence time of logic states. We experimentally verify the enhancement effect on the coherence of qubit motion, and combine it with Carr-Purcell-Meiboom-Gill (CPMG) pulse sequence \cite{PhysRev.94.630,doi:10.1063/1.1716296}, the Motion-CPMG sequence is experimentally demonstrated to provide a stronger protection for logic qubit's coherence.

\section{Principle}
In a multi-qubit quantum system, a logic state encoded with quantum information is initialized on one of the physical qubits, and then the logic state is transferred along different physical qubits one by one. The decoherence time of the logic qubit depends on the decoherence time of each physical qubit and the noise correlation between any two physical qubits.

Consider a system with $n$ independent physical qubits. Each qubit is coupled to a source of Gaussian fluctuations $\xi_i$ ($i=1,2,\dots,n$) of the energy difference between $\left|0 \right\rangle$ and $\left|1 \right\rangle$. Then the Hamiltonian of the system is:
\begin{equation}
\begin{split}
\label{Hamiltonian}
\centering
	H&=-\frac{1}{2}\sum_{i=1}^{n}\sigma_i^z\xi_i(t),\\
	\left\langle \xi_i(0)\xi_j(t)\right\rangle &=\int \frac{d\omega}{2\pi}S_{i,j}(\omega)e^{-i\omega t}.
	\end{split}
\end{equation}
where $\sigma_i^z$ is the $z$ Pauli matrix of the $i$th qubit $Q_i$, the $S_{i}(\omega)\equiv S_{i,i}(\omega)$ is the spectral density of noise $\xi_i(t)$ in $Q_i$, and the $S_{i,j}(\omega)~(i\neq j)$ induces the noise correlation between $Q_i$ and $Q_j$. Assuming that the logic state spends equal amount of time on each physical qubit, and the transfer time is much shorter than the residence time, the previous work \cite{PhysRevLett.116.010501} gives a general formula of the logic qubit's decoherence:
\begin{equation}
\centering
\label{decoherence}
	\frac{1}{\tau_L^2}=\frac{1}{2n^2}\sum_{i,j=1}^{n}\int \frac{d\omega}{2\pi}S_{i,j}(\omega).
\end{equation}
If the logic state stays on a single qubit $Q_i$ all the time, which means $n=1$, the decoherence time of the logic state is
\begin{equation}
\centering
\label{single_decoherence}
	\tau_{L,n=1}=\tau_i=\sqrt{\frac{2}{\int\frac{d\omega}{2\pi}S_{i,i}(\omega)}},
\end{equation}
while if $n\neq 1$, define
\begin{equation}
\centering
\label{definition_Sij}
	\int\frac{d\omega}{2\pi}S_{i,j}(\omega)\equiv r_{i,j}\int\frac{d\omega}{2\pi}\left[\frac{S_{i,i}(\omega)+S_{j,j}(\omega)}{2} \right],
\end{equation}
where $r_{i,j}$ represents the noise correlation between $Q_i$ and $Q_j$, then the decoherence time of the logic qubit with n-qubit motion can be rewritten as
\begin{equation}
\label{decoherence_final}
\centering
	\tau_L=\sqrt{\frac{n^2}{\sum_{i=1}^{n}\tau_i^{-2}+\sum_{i< j}r_{i,j}(\tau_i^{-2}+\tau_j^{-2})}}.
\end{equation}
The noise correlation coefficient $r_{i,j}$ has the property $\left|  r_{ij}\right| < 1$, with $r_{ij}=\pm 1$ corresponding to full correlation and full anticorrelation, respectively. According to [Eq.~(\ref{decoherence_final})], one can determine the noise correlation coefficient between any two physical qubits in the system $r_{i,j~(1\leq i<j\leq n)}$ through a series of measurements of qubit motion.

\section{Experiment and results}
\begin{figure}
	\centering
	\includegraphics{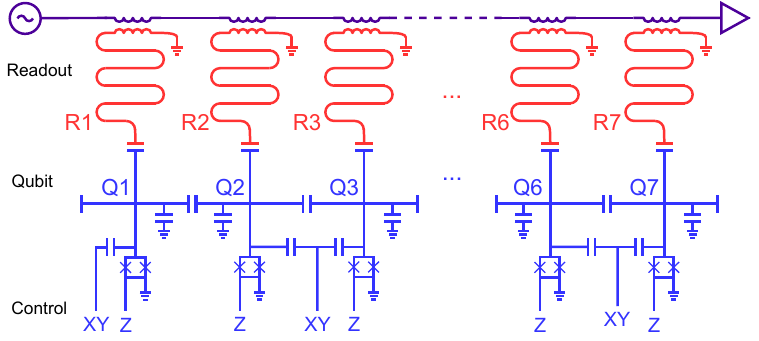}
	\caption{Circuit diagram of the experimental device. Qubits are coupled by capacitance. The transition frequency of each qubit is adjustable. $Q_2$ and $Q_3$, $Q_4$ and $Q_5$, $Q_6$ and $Q_7$ share common XY drive lines.}
	\label{FIG:1}
\end{figure}

The device we use is a chain of seven superconducting Xmon qubits, and the circuit diagram of which is shown in Fig.~\ref{FIG:1}. The adjacent qubits are coupled by capacitance. The transition frequency of each qubit is independently tunable, and independent $XY$ drives can be realized when the frequencies of the qubits do not coincide with each other, although there are some cases that two qubits share one drive line. Seven, four and three adjacent qubits are selected from the same device for the following three experiments respectively.

\begin{figure*}
	\centering
	\includegraphics{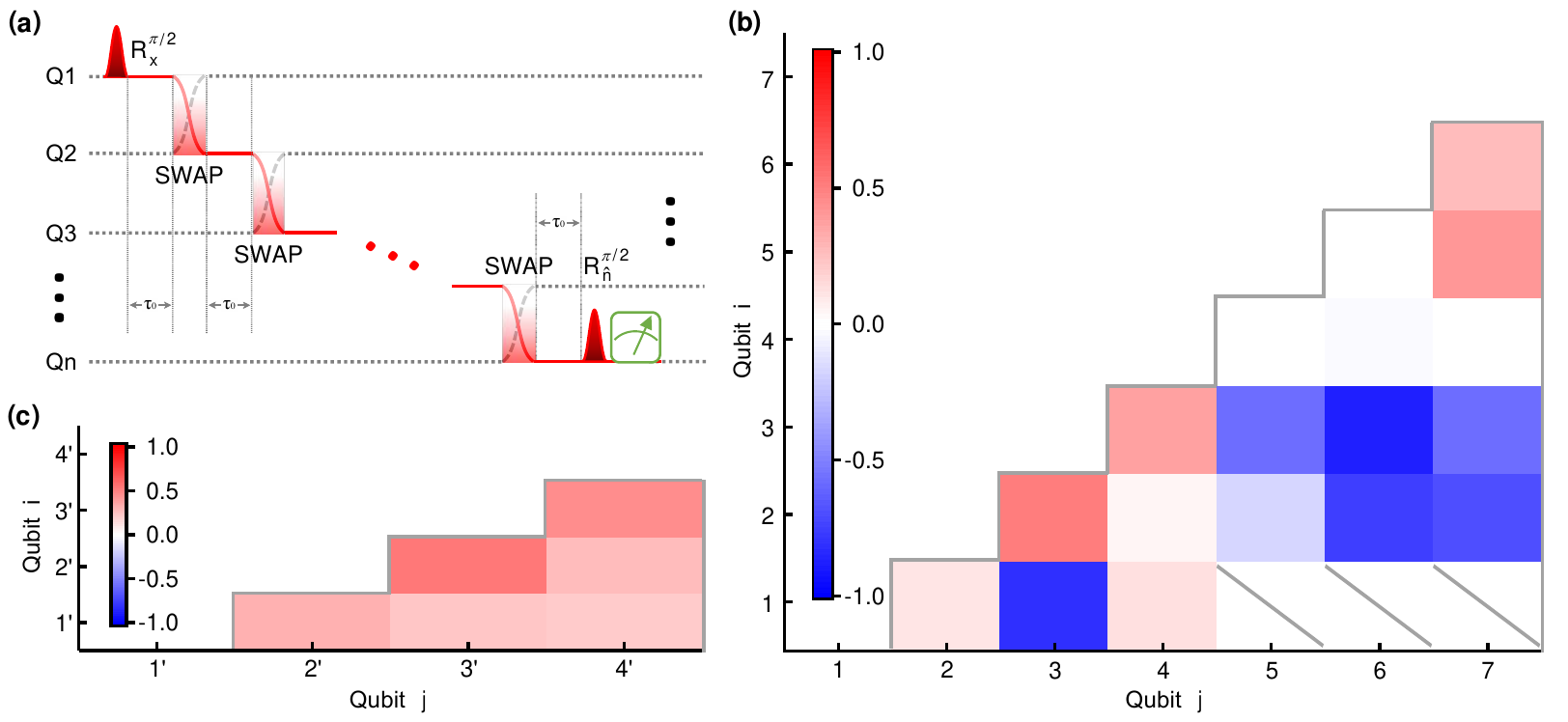}
	\caption{\textbf{(a)} The experimental sequence for measuring the Ramsey fringe of the logic qubit with n-qubit motion. The logic qubit spends equal amount of time $\tau_0$ on each physical qubit. The SWAP gates are realized by adjusting two adjacent qubits to frequency resonance for a short period of time. The rotation operation on the last qubit has a time-dependent direction $ \hat{n}=\left\langle cos(\omega_r t), sin(\omega_r t), 0 \right\rangle$. \textbf{(b)} The measured noise correlation matrix of a 7-qubit system. The gray slash indicates that the corresponding data is indeterminate. \textbf{(c)} The measured $r_{i,j}$ of a subsystem with 4 qubits. $Q_{1^\prime}$, $Q_{2^\prime}$, $Q_{3^\prime}$ and $Q_{4^\prime}$ are the same qubits as $Q_1$, $Q_2$, $Q_3$ and $Q_4$ respectively, but they are at different operating frequencies.}
	\label{FIG:2}
\end{figure*}

The sequence of measuring Ramsey fringe of a logic qubit with n-qubit motion is shown in Fig.~\ref{FIG:2}\textcolor{blue}{(a)} \cite{PhysRevLett.116.010501}. Firstly, the qubits are tuned to their operating frequencies. Then we prepare the initial logic state $\left| \psi_0\right\rangle _L=\left| 0\right\rangle -i\left| 1\right\rangle$ on the first physical qubit. The logic state is transferred successively to the second, the third, $\dots$, until the last physical qubit by a series of SWAP gates, which is realized by adjusting two adjacent qubits to frequency resonance for a short period of time (around 10ns, far shorter than the residence time of logic qubit on each physical qubit). During the process of motion, the residence time of logic state on each physical qubit $\tau_0$ is artificially set to be equal. Finally, the last physical qubit is rotated with a unitary operation $R_{\hat{n}}^{\pi/2}$, where $ \hat{n}=\left\langle cos(\omega_r t), sin(\omega_r t), 0 \right\rangle$ ($\omega_r$ is the rotation frequency of $\hat{n}$), then we measure the population of state $\left| 1\right\rangle$ on the last physical qubit, and thus the Ramsey fringe of the logic state is obtained. Fitting the curves with
\begin{equation}
\label{fitting_equation}
P_{\left|1 \right\rangle }(t) \propto exp\left(-\frac{t}{2T_1^{ave}}-\frac{t^2}{\tau_L^2}\right)~cos(\omega_r t+\phi),
\end{equation}
where $T_1^{ave}$ is the average of the $T_1$ for all the qubits attending the motion, one can get the decoherence time $\tau_L$ of the logic qubit \cite{PhysRevLett.109.067001}.

We determine the noise correlation coefficient $r_{i,j}$ between any two physical qubits in the system by the method of qubit motion. First, we adjust the 7 qubits to their operating frequencies with no interference, and measure their decoherence time $T_2^*$ at their own operating frequencies. Then, we measure the Ramsey fringes of logic states with qubit motion between all groups of two adjacent qubits (e.g., $Q_1 \to Q_2$, $Q_2 \to Q_3$, and so on), and get the decoherence time $\tau_{L,n=2}$ of each set of measurements by fitting the Ramsey fringes with [Eq.~(\ref{fitting_equation})] as mentioned above, and thus $r_{i,j~(j=i+1)}$ could be calculated according to [Eq.~(\ref{decoherence_final})]. Next, we measure the Ramsey fringes of logic states with all groups of three-adjacent-qubit motion ($Q_1 \to Q_2 \to Q_3$, $Q_2 \to Q_3 \to Q_4$, $\dots$). By using the $r_{i,j~(j=i+1)}$ measured previously, $r_{i,j~(j=i+2)}$ could be calculated according to [Eq.~(\ref{decoherence_final})]. The same could be done to all sets of four-, five-, six-, and seven-adjacent-qubit motion. Finally, a total of 28 Ramsey fringes are measured and fitted, until we get $r_{i,j~(j=i+6)}$. By then, we could determine all the noise correlation coefficients between any two physical qubits in the 7-qubit quantum system.

Fig.~\ref{FIG:2}\textcolor{blue}{(b)} shows the measured noise correlation matrix of the seven qubits. As is indicated by [Eq.~(\ref{definition_Sij})], $r_{i,j}=r_{j,i}$, and $r_{i,i}=1$, so we only show the elements $r_{i,j~(i<j)}$ in Fig.~\ref{FIG:2}\textcolor{blue}{(b)}, as well as in Fig.~\ref{FIG:2}\textcolor{blue}{(c)}. The gray slash indicates that the corresponding data is indeterminate. As the experiment is repeated, the measured $r_{1,5}$ fluctuates within a range of almost the same magnitude as itself, so the measured value is not reliable. While $r_{1,6}$ and $r_{1,7}$ need to be calculated from $r_{1,5}$, thus $r_{1,6}$ and $r_{1,7}$ cannot be determined either. Besides the previous measured $r_{i,j~(i<j<5)}$, the accuracy of gates and the stability of the measurement system will also influence the uncertainty of $r_{1,5}$, and the indeterminate $r_{1,5}$ in our experiment is induced by the fact that $Q_1$ and $Q_5$ have non-ignorable driving crosstalk since their detuning is only around 70 MHz. In general, all the $r_{i,j}$ can be measured by the method of qubit motion. Moreover, we can see from Fig.\ref{FIG:2}\textcolor{blue}{(b)} that all the correlation coefficients associated with $Q_3$ are larger in absolute value, which reveals that there is a noise source located near $Q_3$, and it affects all the qubits in the system. Another phenomenon also supports this conclusion: compared with other identical qubits, the decoherence time of $Q_3$ is only about a third of the decoherence time of other qubits under the same magnetic flux. The $r_{2,6}$, $r_{2,7}$, and $r_{6,7}$ have large absolute values simultaneously, which most likely means that the three qubits share a common noise source. And the large noise correlation between $Q_5$ and $Q_7$ is due to the driving crosstalk, as their operating frequencies differ by only about 90 MHz. One can find the detailed information in Table \ref{table_2} in the Appendix. Furthermore, one should note that, for the same physical qubits, the noise correlation coefficients between the two qubits may be different when the qubits are at different operating frequencies, because the  noise is frequency dependent, which is explicit in [Eq.~(\ref{definition_Sij})]. The experimental results (Fig.~\ref{FIG:2}\textcolor{blue}{(c)}) also show the different noise correlations when $Q_1$, $Q_2$, $Q_3$, $Q_4$ are at different operating frequencies (denoted as $Q_{1^\prime}$, $Q_{2^\prime}$, $Q_{3^\prime}$, $Q_{4^\prime}$). The above results show that qubit motion is an efficient method for characterizing spacial noise correlation, and what's even more exhilarating is that, for a multi-qubit quantum system with a structure of two-dimensional array, the steps required to calibrate the noise correlation between any two physical qubits by the method of qubit motion will not increase rapidly with the increase of qubit number, because we can choose the shortest route for the motion of logic state by taking advantage of the coupling structure. Furthermore, the above results indicate that, the characterization of noise correlation between any two qubits provides a judgment basis for the diagnosis of noise, and offers a way to do better correlation-sensitive tasks by selecting appropriate operating frequencies for the physical qubits, for instance, getting a set of operating frequencies for physical qubits to do QEC, at which the physical qubits have weak noise correlations.

\begin{figure}
	\centering
	\includegraphics{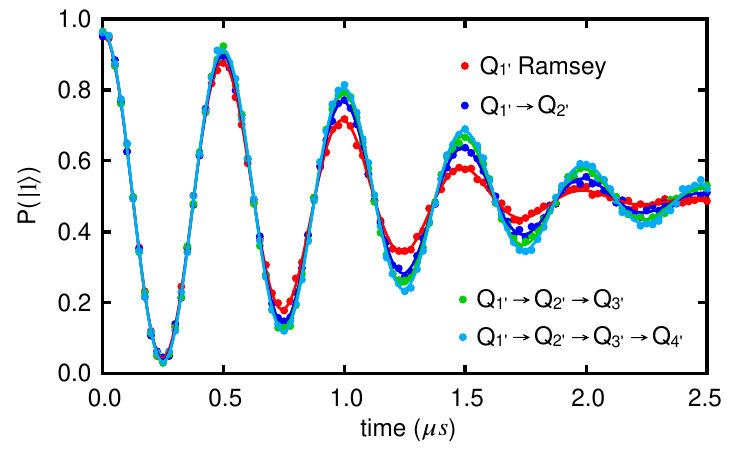}
	\caption{The coherence enhancement of logic qubit with qubit motion. The dots are the experimental data with an average of 10 thousands repeated measurements, and the lines are the fitting curves of the experimental data with [Eq.~(\ref{fitting_equation})]. The $T_2^*$ of physical qubit $Q_{1^\prime}$ is $1.12\mu s$ (Red), while the decoherence time of the logic qubit with 2-, 3-, and 4-qubit motion is extended to $1.41\mu s$ (Blue), $1.55\mu s$ (Green) and $1.65\mu s$ (Cyan) respectively.}
	\label{FIG:3}
\end{figure}

In addition to characterizing the noise correlation between any two physical qubits, qubit motion can also enhance the coherence of logic qubit, which has been observed in a phase-qubit system with all qubits coupled together through a common resonator \cite{PhysRevLett.116.010501}. We have verified this effect in our quantum system, and our results indicate that qubit motion still has suppression effect on decoherence even if the system has a different structure, coupling form and noise level. In order to make the effect easy to be observed, we first adjust the four adjacent physical qubits to the same flux-noise level, which means that they have basically equal $T_2^*$ \cite{PhysRevB.85.174521}, by tuning their transition frequencies. Then the sequence of qubit motion is applied to the four qubits. Fig.~\ref{FIG:3} shows the Ramsey fringes of logic qubit with single-, two-, three- and four-qubit motion. Fitting the Ramsey fringes with [Eq.~(\ref{fitting_equation})], the decoherence time $\tau_L$ of the logic qubit with two-, three- and four-qubit motion are $1.41\mu s$, $1.55\mu s$, and $1.65\mu s$ respectively, which shows a definitive improvement compared with the $T_2^*$ of the single physical qubits. Table~\ref{table_1} lists the detail results of this experiment, from which we find that the qubit motion of arbitrary adjacent physical qubits will enhance the coherence of logic state, and the degree of this enhancement depends on each physical qubit's coherence, the noise correlation between any two physical qubits in the motion, and the number of physical qubits involved in the motion.
\begin{table}[cols=20,pos=h]
	\caption{The detailed information of four-qubit motion}\label{table_1}
	\begin{tabular*}{\tblwidth}{@{} LLLLLLLLLLLLLLLLLLLL@{} }
		\toprule
		\toprule
		\multicolumn{8}{c}{Qubit ID} & \multicolumn{3}{c}{$Q_{1^\prime}$} & \multicolumn{3}{c}{$Q_{2^\prime}$} & \multicolumn{3}{c}{$Q_{3^\prime}$} & \multicolumn{3}{c}{$Q_{4^\prime}$} \\
		\midrule
		\multicolumn{8}{c}{Frequency (GHz)} & \multicolumn{3}{c}{3.690} & \multicolumn{3}{c}{4.350} & \multicolumn{3}{c}{4.291} & \multicolumn{3}{c}{3.824} \\
		\multicolumn{8}{c}{$T_1~(\mu s)$} & \multicolumn{3}{c}{12.9} & \multicolumn{3}{c}{14.0} & \multicolumn{3}{c}{12.8} & \multicolumn{3}{c}{10.4} \\
		\multicolumn{8}{c}{$T_2^*~(\mu s)$} & \multicolumn{3}{c}{1.12} & \multicolumn{3}{c}{1.17} & \multicolumn{3}{c}{1.23} & \multicolumn{3}{c}{1.14} \\
		\midrule
		\midrule
		\multicolumn{8}{c}{qubit motion} & \multicolumn{4}{c}{$\tau_L~(\mu s)$} & \multicolumn{2}{c}{correlation} & \multicolumn{6}{c}{$r_{i,j}(\pm)$} \\
		\midrule
		\multicolumn{8}{c}{$Q_{1^\prime}$$\to$$Q_{2^\prime}$} &\multicolumn{4}{c}{1.41} & \multicolumn{2}{c}{$r_{1^\prime,2^\prime}$} & \multicolumn{6}{c}{0.32 (0.02)}\\
		\multicolumn{8}{c}{$Q_{2^\prime}$$\to$$Q_{3^\prime}$} &\multicolumn{4}{c}{1.37} & \multicolumn{2}{c}{$r_{2^\prime,3^\prime}$} & \multicolumn{6}{c}{0.53 (0.02)}\\
		\multicolumn{8}{c}{$Q_{3^\prime}$$\to$$Q_{4^\prime}$} &\multicolumn{4}{c}{1.39} & \multicolumn{2}{c}{$r_{3^\prime,4^\prime}$} & \multicolumn{6}{c}{0.45 (0.02)}\\
		\multicolumn{8}{c}{$Q_{1^\prime}$$\to$$Q_{2^\prime}$$\to$$Q_{3^\prime}$} &\multicolumn{4}{c}{1.55} & \multicolumn{2}{c}{$r_{1^\prime,3^\prime}$} & \multicolumn{6}{c}{0.23 (0.03)}\\
		\multicolumn{8}{c}{$Q_{2^\prime}$$\to$$Q_{3^\prime}$$\to$$Q_{4^\prime}$} &\multicolumn{4}{c}{1.51} & \multicolumn{2}{c}{$r_{2^\prime,4^\prime}$} & \multicolumn{6}{c}{0.27 (0.04)}\\
		\multicolumn{8}{c}{$Q_{1^\prime}$$\to$$Q_{2^\prime}$$\to$$Q_{3^\prime}$$\to$$Q_{4^\prime}$} &\multicolumn{4}{c}{1.65} & \multicolumn{2}{c}{$r_{1^\prime,4^\prime}$} & \multicolumn{6}{c}{0.20 (0.05)}\\
		\bottomrule
		\bottomrule
	\end{tabular*}
\end{table}


\begin{figure*}
	\centering
	\includegraphics[scale=1]{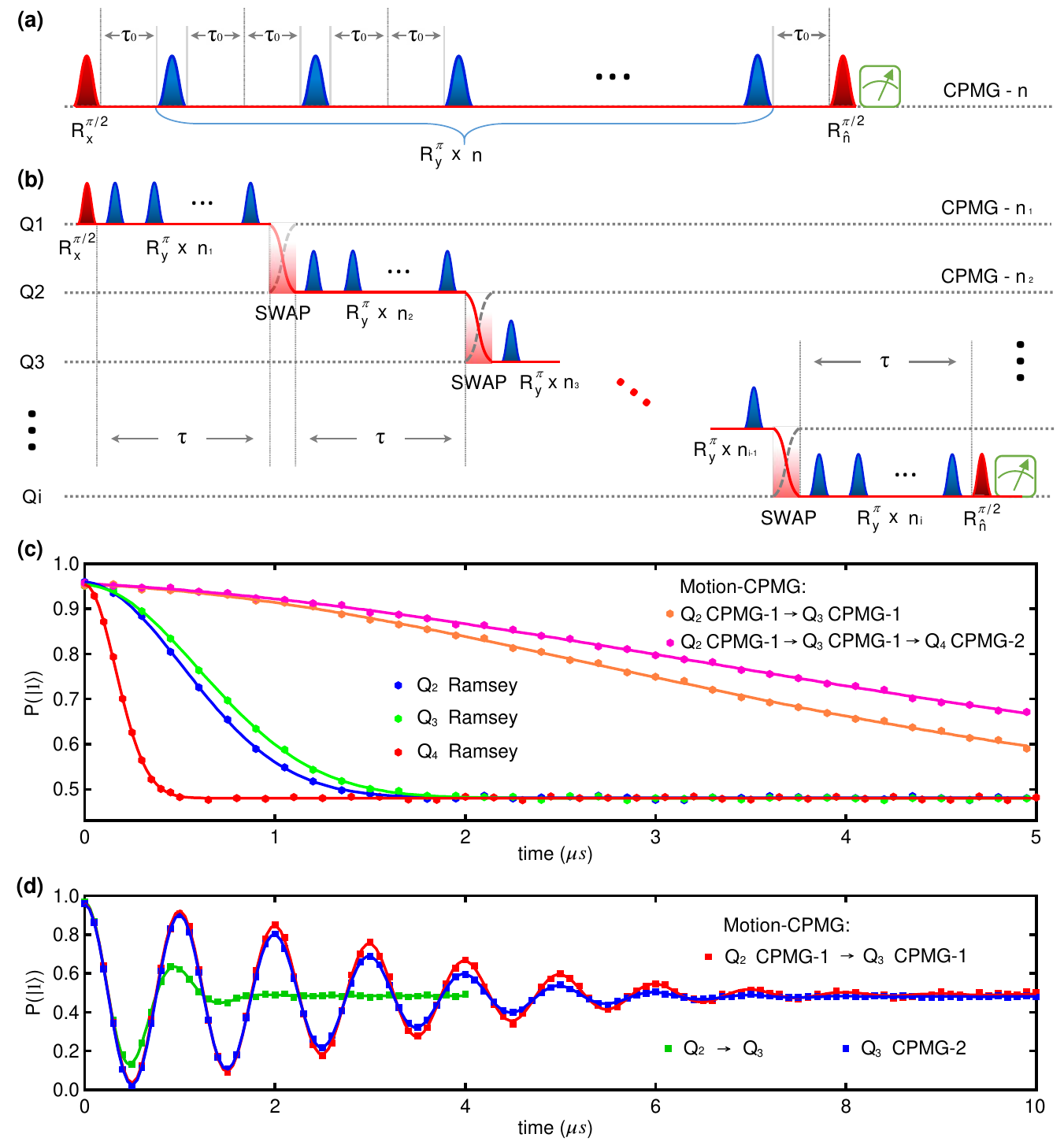}
	\caption{\textbf{(a)} The CPMG-n pulse sequence of a single qubit. \textbf{(b)} The Motion-CPMG operation sequence for an i-qubit system. \textbf{(c)} The coherence enhancement of the logic qubit under an Motion-CPMG operation sequence. The dots are the experimental data with an average of 10 thousands repeatings, and the lines are the fitting curves with the envelope $exp(-t/2T_1^{ave}-t^2/\tau_L^2)$. The $T_2^*$ of $Q_2$, $Q_3$ and $Q_4$ are 754$ns$ (Blue), 872$ns$ (Green), and 233$ns$ (Red), respectively. The $\tau_L$ of the logic qubit under $Q_2$ CPMG-1 $\to Q_3$ CPMG-1 operation sequence and $Q_2$ CPMG-1 $\to Q_3$ CPMG-1 $\to Q_4$ CPMG-2 operation sequence are 4.49$\mu s$ and 5.87$\mu s$ respectively. \textbf{(d)} A comparison of the decoherence of logic qubit between applying Motion-CPMG operation sequence (Red), pure CPMG pulse sequence (Blue) and pure qubit motion sequence (Green). The decoherence times of the logic qubit under Motion-CPMG ($Q_2$ CPMG-1 $\to$ $Q_3$ CPMG-1) sequence, the single physcial qubit $Q_2$ under pure CPMG-2 sequence, $Q_3$ under pure CPMG-2 sequence, and the logic qubit under pure 2-qubit motion sequence ($Q_2 \to Q_3$) are $4.49\mu s$, $3.46\mu s$, $3.61\mu s$ and $0.93\mu s$ respectively. The Ramsey fringe of $Q_2$ under CPMG-2 pulse sequence is not shown in the figure for the recognizability of the figure.}
	\label{FIG:4}
\end{figure*}

The motion of logic state reduces the time that the logic state spends at each local noisy spot, and greatly suppresses the influence of noise distributed at a certain point in the space on dephasing. Therefore, we can say that qubit motion reduces the effect of space-dependent noise. However, for each location where the logic state stays, low-frequency noise causes spectral diffusion of qubit's transition frequency, and results in decoherence. The famous CPMG pulse sequence \cite{PhysRev.94.630,doi:10.1063/1.1716296} can be utilized to effectively suppress this kind of decoherence. So, if we combine qubit motion and CPMG pulse sequence together, we can protect the coherence of the logic state against noise both in time and in space, thus further improving the coherence of the logic qubit.

The schematic diagram of a single-qubit CPMG-n pulse sequence is shown in Fig.~\ref{FIG:4}\textcolor{blue}{(a)}, $n$ equally-spaced $R_y^\pi$ pulses are inserted into the Ramsey sequence, which flip the qubit multiple times to cancel the effect of spectral diffusion on dephasing. The operation sequence that combines qubit motion with CPMG pulse sequence (Motion-CPMG) is shown in Fig.~\ref{FIG:4}\textcolor{blue}{(b)}: based on the sequence of qubit motion, $n_i$ $R_y^\pi$ pulses are inserted into the period when logic state remains on physical qubit Q$_i$. Fig.~\ref{FIG:4}\textcolor{blue}{(c)} shows the experimental results of Motion-CPMG operation sequences, where the rotation before the final measurement is around the direction of $\hat{x}$, so that the Ramsey fringe has only an envelope of $exp(-t/2T_1^{ave}-t^2/\tau_L^2)$ \cite{PhysRevLett.109.067001}. The $T_2^*$ of $Q_2$, $Q_3$, $Q_4$ at their operating frequencies are 754$ns$, 872$ns$, and 233$ns$ respectively, however, when an Motion-CPMG operation sequence is applied to these three physical qubits, the dephasing time of the logic state is extended to 5.87$\mu s$, an increase by nearly an order of magnitude. In order to distinguish the effect of Motion-CPMG operation sequence on the enhancement of coherence from that of pure CPMG sequence and pure qubit motion, several comparisons are made. For the situation of Motion-CPMG ($Q_2$ CPMG-1 $\to$ $Q_3$ CPMG-1), it is equivalent to a logic CPMG-2 pulse sequence for the logic qubit, so we measure the decoherence time of the single physical qubit ($Q_2$ and $Q_3$ respectively) under a pure CPMG-2 pulse sequence as a comparison, and the pure qubit motion $Q_2 \to Q_3$ is measured too. The results are shown in Fig.~\ref{FIG:4}\textcolor{blue}{(d)}, the decoherence times of the logic qubit under Motion-CPMG ($Q_2$ CPMG-1 $\to$ $Q_3$ CPMG-1) sequence, the single physcial qubit $Q_2$ under pure CPMG-2 sequence, the single physcial qubit $Q_3$ under pure CPMG-2 sequence, and the logic qubit under pure 2-qubit motion sequence ($Q_2 \to Q_3$) are $4.49\mu s$, $3.46\mu s$, $3.61\mu s$ and $0.93\mu s$ respectively, which reveals that Motion-CPMG operation sequence suppresses decoherence more significantly. More detailed information can be found in Table \ref{table_3} in the Appendix.

\section{Conclusion}
In summary, we experimentally characterize the noise correlation between any two physical qubits in a multi-qubit system by an efficient method of qubit motion. This relay-based quantum state transfer method can also be applied to the two-dimensional array structure required by the surface code QEC scheme, which paves a way for the realization of QEC and fault-tolerant quantum computation. The measured correlations offer a judgment basis for the diagnosis of noise, and our results show that one can do a better correlation-sensitive task by selecting appropriate operating frequencies for the physical qubits. In addition, we propose an efficacious operation sequence which combines qubit motion and CPMG technique to enhance the coherence of logic qubit, and the experimental results show that the decoherence time of logic qubit is extended by nearly an order of magnitude.
\\
\\
\noindent \textbf{Acknowlegement:}
\\
This work was supported by the NSFC of China (Grants No. 11890704, 12004042, 11674376, 11905100), the NSF of Beijing (Grants No. Z190012), National Key Research and Development Program of China (Grants No. 2016YFA0301800)\\ 
and the Key-Area Research and Development Program of GuangDong Province (Grants No. 2018B030326001). Thanks to \textbf{Yu Song} for her advice on English expression during the drafting of this manuscript.


\bibliographystyle{model3-num-names}
\bibliography{citations}

\begin{thebibliography}{34}
\providecommand{\natexlab}[1]{#1}
\providecommand{\url}[1]{\texttt{#1}}
\providecommand{\href}[2]{#2}
\providecommand{\path}[1]{#1}
\providecommand{\eprint}[1]{\href{http://arxiv.org/abs/#1}{\path{#1}}}
\providecommand{\DOIprefix}{doi:}
\providecommand{\ArXivprefix}{arXiv:}
\providecommand{\URLprefix}{URL: }
\providecommand{\Pubmedprefix}{pmid:}
\providecommand{\doi}[1]{\href{http://dx.doi.org/#1}{\path{#1}}}
\providecommand{\Pubmed}[1]{\href{pmid:#1}{\path{#1}}}
\providecommand{\BIBand}{and}
\providecommand{\bibinfo}[2]{#2}
\ifx\xfnm\undefined \def\xfnm[#1]{\unskip,\space#1}\fi
\bibitem[{Preskill(2018)}]{Preskill_2018}
\bibinfo{author}{Preskill\xfnm[ J.]}.
\newblock \bibinfo{title}{Quantum computing in the nisq era and beyond}.
\newblock \bibinfo{journal}{Quantum}
  \bibinfo{year}{2018};\bibinfo{volume}{2}:\bibinfo{pages}{79}.
\newblock \DOIprefix\doi{10.22331/q-2018-08-06-79}.
\bibitem[{Terhal(2015)}]{Terhal2015Quantum}
\bibinfo{author}{Terhal\xfnm[ B.M.]}.
\newblock \bibinfo{title}{Quantum error correction for quantum memories}.
\newblock \bibinfo{journal}{Rev Mod Phys}
  \bibinfo{year}{2015};\bibinfo{volume}{87}:\bibinfo{pages}{307--346}.
\newblock \DOIprefix\doi{10.1103/RevModPhys.87.307}.
\bibitem[{Gambetta et~al.(2012)Gambetta, C\'orcoles, Merkel, Johnson, Smolin,
  Chow et~al.}]{PhysRevLett.109.240504}
\bibinfo{author}{Gambetta\xfnm[ J.M.]}, \bibinfo{author}{C\'orcoles\xfnm[
  A.D.]}, \bibinfo{author}{Merkel\xfnm[ S.T.]}, \bibinfo{author}{Johnson\xfnm[
  B.R.]}, \bibinfo{author}{Smolin\xfnm[ J.A.]}, \bibinfo{author}{Chow\xfnm[
  J.M.]}, et~al.
\newblock \bibinfo{title}{Characterization of addressability by simultaneous
  randomized benchmarking}.
\newblock \bibinfo{journal}{Phys Rev Lett}
  \bibinfo{year}{2012};\bibinfo{volume}{109}:\bibinfo{pages}{240504}.
\newblock \DOIprefix\doi{10.1103/PhysRevLett.109.240504}.
\bibitem[{Aharonov et~al.(2006)Aharonov, Kitaev and
  Preskill}]{PhysRevLett.96.050504}
\bibinfo{author}{Aharonov\xfnm[ D.]}, \bibinfo{author}{Kitaev\xfnm[ A.]},
  \bibinfo{author}{Preskill\xfnm[ J.]}.
\newblock \bibinfo{title}{Fault-tolerant quantum computation with long-range
  correlated noise}.
\newblock \bibinfo{journal}{Phys Rev Lett}
  \bibinfo{year}{2006};\bibinfo{volume}{96}:\bibinfo{pages}{050504}.
\newblock \DOIprefix\doi{10.1103/PhysRevLett.96.050504}.
\bibitem[{Chubb and Flammia(2018)}]{chubb2018statistical}
\bibinfo{author}{Chubb\xfnm[ C.T.]}, \bibinfo{author}{Flammia\xfnm[ S.T.]}.
\newblock \bibinfo{title}{Statistical mechanical models for quantum codes with
  correlated noise}.
\newblock \bibinfo{year}{2018}.
\newblock \href{http://arxiv.org/abs/1809.10704}{\tt arXiv:1809.10704}.
\bibitem[{Aliferis et~al.(2005)Aliferis, Gottesman and
  Preskill}]{aliferis2005quantum}
\bibinfo{author}{Aliferis\xfnm[ P.]}, \bibinfo{author}{Gottesman\xfnm[ D.]},
  \bibinfo{author}{Preskill\xfnm[ J.]}.
\newblock \bibinfo{title}{Quantum accuracy threshold for concatenated
  distance-3 codes}.
\newblock \bibinfo{year}{2005}.
\newblock \href{http://arxiv.org/abs/quant-ph/0504218}{\tt
  arXiv:quant-ph/0504218}.
\bibitem[{{Shor}(1996)}]{548464}
\bibinfo{author}{{Shor}\xfnm[ P.W.]}.
\newblock \bibinfo{title}{Fault-tolerant quantum computation}.
\newblock In: \bibinfo{booktitle}{Proceedings of 37th Conference on Foundations
  of Computer Science}. \bibinfo{year}{1996}, p. \bibinfo{pages}{56--65}.
\newblock \DOIprefix\doi{10.1109/SFCS.1996.548464}.
\bibitem[{Gottesman(1997)}]{gottesman1997stabilizer}
\bibinfo{author}{Gottesman\xfnm[ D.]}.
\newblock \bibinfo{title}{Stabilizer codes and quantum error correction}.
\newblock \bibinfo{year}{1997}.
\newblock \href{http://arxiv.org/abs/quant-ph/9705052}{\tt
  arXiv:quant-ph/9705052}.
\bibitem[{Bravyi and Kitaev(1998)}]{bravyi1998quantum}
\bibinfo{author}{Bravyi\xfnm[ S.B.]}, \bibinfo{author}{Kitaev\xfnm[ A.Y.]}.
\newblock \bibinfo{title}{Quantum codes on a lattice with boundary}.
\newblock \bibinfo{year}{1998}.
\newblock \href{http://arxiv.org/abs/quant-ph/9811052}{\tt
  arXiv:quant-ph/9811052}.
\bibitem[{Kitaev(2003)}]{KITAEV20032}
\bibinfo{author}{Kitaev\xfnm[ A.]}.
\newblock \bibinfo{title}{Fault-tolerant quantum computation by anyons}.
\newblock \bibinfo{journal}{Annals of Physics}
  \bibinfo{year}{2003};\bibinfo{volume}{303}(\bibinfo{number}{1}):\bibinfo{pages}{2
  -- 30}.
\newblock \DOIprefix\doi{https://doi.org/10.1016/S0003-4916(02)00018-0}.
\bibitem[{Fowler et~al.(2009)Fowler, Stephens and
  Groszkowski}]{PhysRevA.80.052312}
\bibinfo{author}{Fowler\xfnm[ A.G.]}, \bibinfo{author}{Stephens\xfnm[ A.M.]},
  \bibinfo{author}{Groszkowski\xfnm[ P.]}.
\newblock \bibinfo{title}{High-threshold universal quantum computation on the
  surface code}.
\newblock \bibinfo{journal}{Phys Rev A}
  \bibinfo{year}{2009};\bibinfo{volume}{80}:\bibinfo{pages}{052312}.
\newblock \DOIprefix\doi{10.1103/PhysRevA.80.052312}.
\bibitem[{Bylander et~al.(2011)Bylander, Gustavsson, Yan, Yoshihara, Harrabi,
  Fitch et~al.}]{bylander_noise_2011}
\bibinfo{author}{Bylander\xfnm[ J.]}, \bibinfo{author}{Gustavsson\xfnm[ S.]},
  \bibinfo{author}{Yan\xfnm[ F.]}, \bibinfo{author}{Yoshihara\xfnm[ F.]},
  \bibinfo{author}{Harrabi\xfnm[ K.]}, \bibinfo{author}{Fitch\xfnm[ G.]},
  et~al.
\newblock \bibinfo{title}{Noise spectroscopy through dynamical decoupling with
  a superconducting flux qubit}.
\newblock \bibinfo{journal}{Nature Physics}
  \bibinfo{year}{2011};\bibinfo{volume}{7}(\bibinfo{number}{7}):\bibinfo{pages}{565--570}.
\newblock \DOIprefix\doi{10.1038/nphys1994}.
\bibitem[{Cai et~al.(2012)Cai, Naydenov, Pfeiffer, Mcguinness, Jahnke, Jelezko
  et~al.}]{cai_robust_2012}
\bibinfo{author}{Cai\xfnm[ J.M.]}, \bibinfo{author}{Naydenov\xfnm[ B.]},
  \bibinfo{author}{Pfeiffer\xfnm[ R.]}, \bibinfo{author}{Mcguinness\xfnm[
  L.P.]}, \bibinfo{author}{Jahnke\xfnm[ K.D.]}, \bibinfo{author}{Jelezko\xfnm[
  F.]}, et~al.
\newblock \bibinfo{title}{Robust dynamical decoupling with concatenated
  continuous driving}.
\newblock \bibinfo{journal}{New Journal of Physics}
  \bibinfo{year}{2012};\bibinfo{volume}{14}(\bibinfo{number}{11}):\bibinfo{pages}{113023--113038}.
\newblock \DOIprefix\doi{10.1088/1367-2630/14/11/113023}.
\bibitem[{Wang et~al.(2020)Wang, Liu and Cappellaro}]{wang_coherence_2020}
\bibinfo{author}{Wang\xfnm[ G.]}, \bibinfo{author}{Liu\xfnm[ Y.X.]},
  \bibinfo{author}{Cappellaro\xfnm[ P.]}.
\newblock \bibinfo{title}{Coherence protection and decay mechanism in qubit
  ensembles under concatenated continuous driving}.
\newblock \bibinfo{year}{2020}.
\newblock \href{http://arxiv.org/abs/2008.09027}{\tt arXiv:2008.09027}.
\bibitem[{Yi-Min et~al.(2009)Yi-Min, Yan-Li, Lin-Mei and
  Cheng-Zu}]{Yi_Min_2009}
\bibinfo{author}{Yi-Min\xfnm[ W.]}, \bibinfo{author}{Yan-Li\xfnm[ Z.]},
  \bibinfo{author}{Lin-Mei\xfnm[ L.]}, \bibinfo{author}{Cheng-Zu\xfnm[ L.]}.
\newblock \bibinfo{title}{Quantum gate operations in decoherence-free subspace
  with superconducting charge qubits inside a cavity}.
\newblock \bibinfo{journal}{Chinese Physics Letters}
  \bibinfo{year}{2009};\bibinfo{volume}{26}(\bibinfo{number}{10}):\bibinfo{pages}{100304}.
\newblock \DOIprefix\doi{10.1088/0256-307x/26/10/100304}.
\bibitem[{Lidar et~al.(2001)Lidar, Bacon, Kempe and
  Whaley}]{PhysRevA.63.022306}
\bibinfo{author}{Lidar\xfnm[ D.A.]}, \bibinfo{author}{Bacon\xfnm[ D.]},
  \bibinfo{author}{Kempe\xfnm[ J.]}, \bibinfo{author}{Whaley\xfnm[ K.B.]}.
\newblock \bibinfo{title}{Decoherence-free subspaces for multiple-qubit errors.
  i. characterization}.
\newblock \bibinfo{journal}{Phys Rev A}
  \bibinfo{year}{2001};\bibinfo{volume}{63}:\bibinfo{pages}{022306}.
\newblock \DOIprefix\doi{10.1103/PhysRevA.63.022306}.
\bibitem[{Friesen et~al.(2017)Friesen, Ghosh, Eriksson and
  Coppersmith}]{friesen_decoherence-free_2017}
\bibinfo{author}{Friesen\xfnm[ M.]}, \bibinfo{author}{Ghosh\xfnm[ J.]},
  \bibinfo{author}{Eriksson\xfnm[ M.A.]}, \bibinfo{author}{Coppersmith\xfnm[
  S.N.]}.
\newblock \bibinfo{title}{A decoherence-free subspace in a charge quadrupole
  qubit}.
\newblock \bibinfo{journal}{Nature Communications}
  \bibinfo{year}{2017};\bibinfo{volume}{8}(\bibinfo{number}{1}):\bibinfo{pages}{15923}.
\newblock \DOIprefix\doi{10.1038/ncomms15923}.
\bibitem[{Altepeter et~al.(2004)Altepeter, Hadley, Wendelken, Berglund and
  Kwiat}]{altepeter_experimental_2004}
\bibinfo{author}{Altepeter\xfnm[ J.B.]}, \bibinfo{author}{Hadley\xfnm[ P.G.]},
  \bibinfo{author}{Wendelken\xfnm[ S.M.]}, \bibinfo{author}{Berglund\xfnm[
  A.J.]}, \bibinfo{author}{Kwiat\xfnm[ P.G.]}.
\newblock \bibinfo{title}{Experimental {Investigation} of a {Two}-{Qubit}
  {Decoherence}-{Free} {Subspace}}.
\newblock \bibinfo{journal}{Physical Review Letters}
  \bibinfo{year}{2004};\bibinfo{volume}{92}(\bibinfo{number}{14}):\bibinfo{pages}{147901}.
\newblock \DOIprefix\doi{10.1103/PhysRevLett.92.147901}.
\bibitem[{Averin et~al.(2016{\natexlab{a}})Averin, Xu, Zhong, Song, Wang and
  Han}]{averin_suppression_2016}
\bibinfo{author}{Averin\xfnm[ D.]}, \bibinfo{author}{Xu\xfnm[ K.]},
  \bibinfo{author}{Zhong\xfnm[ Y.]}, \bibinfo{author}{Song\xfnm[ C.]},
  \bibinfo{author}{Wang\xfnm[ H.]}, \bibinfo{author}{Han\xfnm[ S.]}.
\newblock \bibinfo{title}{Suppression of {Dephasing} by {Qubit} {Motion} in
  {Superconducting} {Circuits}}.
\newblock \bibinfo{journal}{Physical Review Letters}
  \bibinfo{year}{2016}{\natexlab{a}};\bibinfo{volume}{116}(\bibinfo{number}{1}):\bibinfo{pages}{010501}.
\newblock \DOIprefix\doi{10.1103/PhysRevLett.116.010501}.
\bibitem[{Karimi and Pekola(2017)}]{PhysRevB.96.115408}
\bibinfo{author}{Karimi\xfnm[ B.]}, \bibinfo{author}{Pekola\xfnm[ J.P.]}.
\newblock \bibinfo{title}{Correlated versus uncorrelated noise acting on a
  quantum refrigerator}.
\newblock \bibinfo{journal}{Phys Rev B}
  \bibinfo{year}{2017};\bibinfo{volume}{96}:\bibinfo{pages}{115408}.
\newblock \DOIprefix\doi{10.1103/PhysRevB.96.115408}.
\bibitem[{Boter et~al.(2020)Boter, Xue, Kr\"ahenmann, Watson, Premakumar, Ward
  et~al.}]{PhysRevB.101.235133}
\bibinfo{author}{Boter\xfnm[ J.M.]}, \bibinfo{author}{Xue\xfnm[ X.]},
  \bibinfo{author}{Kr\"ahenmann\xfnm[ T.]}, \bibinfo{author}{Watson\xfnm[
  T.F.]}, \bibinfo{author}{Premakumar\xfnm[ V.N.]}, \bibinfo{author}{Ward\xfnm[
  D.R.]}, et~al.
\newblock \bibinfo{title}{Spatial noise correlations in a si/sige two-qubit
  device from bell state coherences}.
\newblock \bibinfo{journal}{Phys Rev B}
  \bibinfo{year}{2020};\bibinfo{volume}{101}:\bibinfo{pages}{235133}.
\newblock \DOIprefix\doi{10.1103/PhysRevB.101.235133}.
\bibitem[{Brox et~al.(2012)Brox, Bergli and Galperin}]{brox_bloch-sphere_2012}
\bibinfo{author}{Brox\xfnm[ H.]}, \bibinfo{author}{Bergli\xfnm[ J.]},
  \bibinfo{author}{Galperin\xfnm[ Y.M.]}.
\newblock \bibinfo{title}{Bloch-sphere approach to correlated noise in coupled
  qubits}.
\newblock \bibinfo{journal}{Journal of Physics A: Mathematical and Theoretical}
  \bibinfo{year}{2012};\bibinfo{volume}{45}(\bibinfo{number}{45}):\bibinfo{pages}{455302}.
\newblock \DOIprefix\doi{10.1088/1751-8113/45/45/455302};
  \bibinfo{note}{publisher: IOP Publishing}.
\bibitem[{Villar and Lombardo(2015)}]{villar_decoherence_2015}
\bibinfo{author}{Villar\xfnm[ P.I.]}, \bibinfo{author}{Lombardo\xfnm[ F.C.]}.
\newblock \bibinfo{title}{Decoherence of a solid-state qubit by different noise
  correlation spectra}.
\newblock \bibinfo{journal}{Physics Letters A}
  \bibinfo{year}{2015};\bibinfo{volume}{379}(\bibinfo{number}{4}):\bibinfo{pages}{246--254}.
\newblock \DOIprefix\doi{10.1016/j.physleta.2014.11.022}.
\bibitem[{Kou et~al.(2018)Kou, Smith, Vool, Pop, Sliwa, Hatridge
  et~al.}]{PhysRevApplied.9.064022}
\bibinfo{author}{Kou\xfnm[ A.]}, \bibinfo{author}{Smith\xfnm[ W.C.]},
  \bibinfo{author}{Vool\xfnm[ U.]}, \bibinfo{author}{Pop\xfnm[ I.M.]},
  \bibinfo{author}{Sliwa\xfnm[ K.M.]}, \bibinfo{author}{Hatridge\xfnm[ M.]},
  et~al.
\newblock \bibinfo{title}{Simultaneous monitoring of fluxonium qubits in a
  waveguide}.
\newblock \bibinfo{journal}{Phys Rev Applied}
  \bibinfo{year}{2018};\bibinfo{volume}{9}:\bibinfo{pages}{064022}.
\newblock \DOIprefix\doi{10.1103/PhysRevApplied.9.064022}.
\bibitem[{Schl\"or et~al.(2019)Schl\"or, Lisenfeld, M\"uller, Bilmes,
  Schneider, Pappas et~al.}]{PhysRevLett.123.190502}
\bibinfo{author}{Schl\"or\xfnm[ S.]}, \bibinfo{author}{Lisenfeld\xfnm[ J.]},
  \bibinfo{author}{M\"uller\xfnm[ C.]}, \bibinfo{author}{Bilmes\xfnm[ A.]},
  \bibinfo{author}{Schneider\xfnm[ A.]}, \bibinfo{author}{Pappas\xfnm[ D.P.]},
  et~al.
\newblock \bibinfo{title}{Correlating decoherence in transmon qubits: Low
  frequency noise by single fluctuators}.
\newblock \bibinfo{journal}{Phys Rev Lett}
  \bibinfo{year}{2019};\bibinfo{volume}{123}:\bibinfo{pages}{190502}.
\newblock \DOIprefix\doi{10.1103/PhysRevLett.123.190502}.
\bibitem[{Gustavsson et~al.(2011)Gustavsson, Bylander, Yan, Oliver, Yoshihara
  and Nakamura}]{PhysRevB.84.014525}
\bibinfo{author}{Gustavsson\xfnm[ S.]}, \bibinfo{author}{Bylander\xfnm[ J.]},
  \bibinfo{author}{Yan\xfnm[ F.]}, \bibinfo{author}{Oliver\xfnm[ W.D.]},
  \bibinfo{author}{Yoshihara\xfnm[ F.]}, \bibinfo{author}{Nakamura\xfnm[ Y.]}.
\newblock \bibinfo{title}{Noise correlations in a flux qubit with tunable
  tunnel coupling}.
\newblock \bibinfo{journal}{Phys Rev B}
  \bibinfo{year}{2011};\bibinfo{volume}{84}:\bibinfo{pages}{014525}.
\newblock \DOIprefix\doi{10.1103/PhysRevB.84.014525}.
\bibitem[{von Lüpke et~al.(2019)von Lüpke, Beaudoin, Norris, Sung, Winik, Qiu
  et~al.}]{lpke2019twoqubit}
\bibinfo{author}{von Lüpke\xfnm[ U.]}, \bibinfo{author}{Beaudoin\xfnm[ F.]},
  \bibinfo{author}{Norris\xfnm[ L.M.]}, \bibinfo{author}{Sung\xfnm[ Y.]},
  \bibinfo{author}{Winik\xfnm[ R.]}, \bibinfo{author}{Qiu\xfnm[ J.Y.]}, et~al.
\newblock \bibinfo{title}{Two-qubit spectroscopy of spatiotemporally correlated
  quantum noise in superconducting qubits}.
\newblock \bibinfo{year}{2019}.
\newblock \href{http://arxiv.org/abs/1912.04982}{\tt arXiv:1912.04982}.
\bibitem[{Xu et~al.(2013)Xu, Sun, Li, Xu, Guo, Andersson
  et~al.}]{xu_experimental_2013}
\bibinfo{author}{Xu\xfnm[ J.S.]}, \bibinfo{author}{Sun\xfnm[ K.]},
  \bibinfo{author}{Li\xfnm[ C.F.]}, \bibinfo{author}{Xu\xfnm[ X.Y.]},
  \bibinfo{author}{Guo\xfnm[ G.C.]}, \bibinfo{author}{Andersson\xfnm[ E.]},
  et~al.
\newblock \bibinfo{title}{Experimental recovery of quantum correlations in
  absence of system-environment back-action}.
\newblock \bibinfo{journal}{Nature Communications}
  \bibinfo{year}{2013};\bibinfo{volume}{4}(\bibinfo{number}{1}):\bibinfo{pages}{2851}.
\newblock \DOIprefix\doi{10.1038/ncomms3851}.
\bibitem[{Harper et~al.(2020)Harper, Flammia and
  Wallman}]{harper_efficient_2020}
\bibinfo{author}{Harper\xfnm[ R.]}, \bibinfo{author}{Flammia\xfnm[ S.T.]},
  \bibinfo{author}{Wallman\xfnm[ J.J.]}.
\newblock \bibinfo{title}{Efficient learning of quantum noise}.
\newblock \bibinfo{journal}{Nature Physics}
  \bibinfo{year}{2020};\DOIprefix\doi{10.1038/s41567-020-0992-8}.
\bibitem[{Averin et~al.(2016{\natexlab{b}})Averin, Xu, Zhong, Song, Wang and
  Han}]{PhysRevLett.116.010501}
\bibinfo{author}{Averin\xfnm[ D.V.]}, \bibinfo{author}{Xu\xfnm[ K.]},
  \bibinfo{author}{Zhong\xfnm[ Y.P.]}, \bibinfo{author}{Song\xfnm[ C.]},
  \bibinfo{author}{Wang\xfnm[ H.]}, \bibinfo{author}{Han\xfnm[ S.]}.
\newblock \bibinfo{title}{Suppression of dephasing by qubit motion in
  superconducting circuits}.
\newblock \bibinfo{journal}{Phys Rev Lett}
  \bibinfo{year}{2016}{\natexlab{b}};\bibinfo{volume}{116}:\bibinfo{pages}{010501}.
\newblock \DOIprefix\doi{10.1103/PhysRevLett.116.010501}.
\bibitem[{Carr and Purcell(1954)}]{PhysRev.94.630}
\bibinfo{author}{Carr\xfnm[ H.Y.]}, \bibinfo{author}{Purcell\xfnm[ E.M.]}.
\newblock \bibinfo{title}{Effects of diffusion on free precession in nuclear
  magnetic resonance experiments}.
\newblock \bibinfo{journal}{Phys Rev}
  \bibinfo{year}{1954};\bibinfo{volume}{94}:\bibinfo{pages}{630--638}.
\newblock \DOIprefix\doi{10.1103/PhysRev.94.630}.
\bibitem[{Meiboom and Gill(1958)}]{doi:10.1063/1.1716296}
\bibinfo{author}{Meiboom\xfnm[ S.]}, \bibinfo{author}{Gill\xfnm[ D.]}.
\newblock \bibinfo{title}{Modified spin‐echo method for measuring nuclear
  relaxation times}.
\newblock \bibinfo{journal}{Review of Scientific Instruments}
  \bibinfo{year}{1958};\bibinfo{volume}{29}(\bibinfo{number}{8}):\bibinfo{pages}{688--691}.
\newblock \DOIprefix\doi{10.1063/1.1716296}.
\bibitem[{Sank et~al.(2012)Sank, Barends, Bialczak, Chen, Kelly, Lenander
  et~al.}]{PhysRevLett.109.067001}
\bibinfo{author}{Sank\xfnm[ D.]}, \bibinfo{author}{Barends\xfnm[ R.]},
  \bibinfo{author}{Bialczak\xfnm[ R.C.]}, \bibinfo{author}{Chen\xfnm[ Y.]},
  \bibinfo{author}{Kelly\xfnm[ J.]}, \bibinfo{author}{Lenander\xfnm[ M.]},
  et~al.
\newblock \bibinfo{title}{Flux noise probed with real time qubit tomography in
  a josephson phase qubit}.
\newblock \bibinfo{journal}{Phys Rev Lett}
  \bibinfo{year}{2012};\bibinfo{volume}{109}:\bibinfo{pages}{067001}.
\newblock \DOIprefix\doi{10.1103/PhysRevLett.109.067001}.
\bibitem[{Yan et~al.(2012)Yan, Bylander, Gustavsson, Yoshihara, Harrabi, Cory
  et~al.}]{PhysRevB.85.174521}
\bibinfo{author}{Yan\xfnm[ F.]}, \bibinfo{author}{Bylander\xfnm[ J.]},
  \bibinfo{author}{Gustavsson\xfnm[ S.]}, \bibinfo{author}{Yoshihara\xfnm[
  F.]}, \bibinfo{author}{Harrabi\xfnm[ K.]}, \bibinfo{author}{Cory\xfnm[
  D.G.]}, et~al.
\newblock \bibinfo{title}{Spectroscopy of low-frequency noise and its
  temperature dependence in a superconducting qubit}.
\newblock \bibinfo{journal}{Phys Rev B}
  \bibinfo{year}{2012};\bibinfo{volume}{85}:\bibinfo{pages}{174521}.
\newblock \DOIprefix\doi{10.1103/PhysRevB.85.174521}.

\end{thebibliography}

\appendix
\section{Appendix}

\begin{table*}[cols=26,pos=h,width=.9\textwidth]
	\caption{The detailed information of the 7-qubit motion}\label{table_2}
	\begin{tabular*}{.9\textwidth}{@{} LLLLLLLLLLLLLLLLLLLLLLLLLL@{} }
		\toprule
		\toprule
		\multicolumn{12}{c}{Qubit ID} & \multicolumn{2}{c}{$Q_1$} & \multicolumn{2}{c}{$Q_2$} & \multicolumn{2}{c}{$Q_3$} & \multicolumn{2}{c}{$Q_4$} & \multicolumn{2}{c}{$Q_5$} & \multicolumn{2}{c}{$Q_6$} & \multicolumn{2}{c}{$Q_7$} \\
		\midrule
		\multicolumn{12}{c}{Frequency (GHz)} & \multicolumn{2}{c}{4.276} &
		\multicolumn{2}{c}{3.708} &
		\multicolumn{2}{c}{4.114} &
		\multicolumn{2}{c}{3.607} &
		\multicolumn{2}{c}{4.205} &
		\multicolumn{2}{c}{3.503} &
		\multicolumn{2}{c}{4.294}  \\
		\multicolumn{12}{c}{$T_1~(\mu s)$} & \multicolumn{2}{c}{15.3} &
		\multicolumn{2}{c}{12.8} &
		\multicolumn{2}{c}{11.6} &
		\multicolumn{2}{c}{10.3} &
		\multicolumn{2}{c}{18.6} &
		\multicolumn{2}{c}{19.0} &
		\multicolumn{2}{c}{13.7}  \\
		\multicolumn{12}{c}{$T_2^*~(\mu s)$} & \multicolumn{2}{c}{3.45} &
		\multicolumn{2}{c}{0.75} &
		\multicolumn{2}{c}{0.87} &
		\multicolumn{2}{c}{0.23} &
		\multicolumn{2}{c}{2.35} &
		\multicolumn{2}{c}{1.17} &
		\multicolumn{2}{c}{3.04} \\
		\midrule
		\midrule
		\multicolumn{12}{c}{qubit motion} & \multicolumn{4}{c}{$\tau_L~(\mu s)$} & \multicolumn{4}{c}{correlation} & \multicolumn{6}{c}{$r_{i,j}(\pm)$} \\
		\midrule
		\multicolumn{12}{c}{$Q_1 \to Q_2$} &
		\multicolumn{4}{c}{1.39} &
		\multicolumn{4}{c}{$r_{1,2}$} &
		\multicolumn{6}{c}{0.11 (0.02)} \\
		\multicolumn{12}{c}{$Q_2 \to Q_3$} &
		\multicolumn{4}{c}{0.93} &
		\multicolumn{4}{c}{$r_{2,3}$} &
		\multicolumn{6}{c}{0.51 (0.03)} \\
		\multicolumn{12}{c}{$Q_3 \to Q_4$} &
		\multicolumn{4}{c}{0.38} &
		\multicolumn{4}{c}{$r_{3,4}$} &
		\multicolumn{6}{c}{0.37 (0.02)} \\
		\multicolumn{12}{c}{$Q_4 \to Q_5$} &
		\multicolumn{4}{c}{0.46} &
		\multicolumn{4}{c}{$r_{4,5}$} &
		\multicolumn{6}{c}{-0.01 (0.01)} \\
		\multicolumn{12}{c}{$Q_5 \to Q_6$} &
		\multicolumn{4}{c}{2.09} &
		\multicolumn{4}{c}{$r_{5,6}$} &
		\multicolumn{6}{c}{0.00 (0.02)} \\
		\multicolumn{12}{c}{$Q_6 \to Q_7$} &
		\multicolumn{4}{c}{1.94} &
		\multicolumn{4}{c}{$r_{6,7}$} &
		\multicolumn{6}{c}{0.27 (0.03)} \\
		\midrule
		\multicolumn{12    }{c}{$Q_1 \to Q_2 \to Q_3$} &
		\multicolumn{4}{c}{1.54} &
		\multicolumn{4}{c}{$r_{1,3}$} &
		\multicolumn{6}{c}{-0.82 (0.04)} \\
		\multicolumn{12}{c}{$Q_2 \to Q_3 \to Q_4$} &
		\multicolumn{4}{c}{0.53} &
		\multicolumn{4}{c}{$r_{2,4}$} &
		\multicolumn{6}{c}{0.05 (0.03)} \\
		\multicolumn{12}{c}{$Q_3 \to Q_4 \to Q_5$} &
		\multicolumn{4}{c}{0.64} &
		\multicolumn{4}{c}{$r_{3,5}$} &
		\multicolumn{6}{c}{-0.57 (0.13)} \\
		\multicolumn{12}{c}{$Q_4 \to Q_5 \to Q_6$} &
		\multicolumn{4}{c}{0.69} &
		\multicolumn{4}{c}{$r_{4,6}$} &
		\multicolumn{6}{c}{-0.04 (0.02)} \\
		\multicolumn{12}{c}{$Q_5 \to Q_6 \to Q_7$} &
		\multicolumn{4}{c}{2.56} &
		\multicolumn{4}{c}{$r_{5,7}$} &
		\multicolumn{6}{c}{0.42 (0.11)} \\
		\midrule
		\multicolumn{12}{c}{$Q_1 \to Q_2 \to Q_3 \to Q_4$} &
		\multicolumn{4}{c}{0.69} &
		\multicolumn{4}{c}{$r_{1,4}$} &
		\multicolumn{6}{c}{0.13 (0.07)} \\
		\multicolumn{12}{c}{$Q_2 \to Q_3 \to Q_4 \to Q_5$} &
		\multicolumn{4}{c}{0.72} &
		\multicolumn{4}{c}{$r_{2,5}$} &
		\multicolumn{6}{c}{-0.17 (0.14)} \\
		\multicolumn{12}{c}{$Q_3 \to Q_4 \to Q_5 \to Q_6$} &
		\multicolumn{4}{c}{0.80} &
		\multicolumn{4}{c}{$r_{3,6}$} &
		\multicolumn{6}{c}{-0.87 (0.16)} \\
		\multicolumn{12}{c}{$Q_4 \to Q_5 \to Q_6 \to Q_7$} &
		\multicolumn{4}{c}{0.90} &
		\multicolumn{4}{c}{$r_{4,7}$} &
		\multicolumn{6}{c}{0.02 (0.01)} \\
		\midrule
		\multicolumn{12}{c}{$Q_1 \to Q_2 \to Q_3 \to Q_4 \to Q_5$} &
		\multicolumn{4}{c}{0.84} &
		\multicolumn{4}{c}{$r_{1,5}$} &
		\multicolumn{6}{c}{--} \\
		\multicolumn{12}{c}{$Q_2 \to Q_3 \to Q_4 \to Q_5 \to Q_6$} &
		\multicolumn{4}{c}{0.85} &
		\multicolumn{4}{c}{$r_{2,6}$} &
		\multicolumn{6}{c}{-0.75 (0.17)} \\
		\multicolumn{12}{c}{$Q_3 \to Q_4 \to Q_5 \to Q_6 \to Q_7$} &
		\multicolumn{4}{c}{0.89} &
		\multicolumn{4}{c}{$r_{3,7}$} &
		\multicolumn{6}{c}{-0.57 (0.22)} \\
		\midrule
		\multicolumn{12}{c}{$Q_1 \to Q_2 \to Q_3 \to Q_4 \to Q_5 \to Q_6$} &
		\multicolumn{4}{c}{1.20} &
		\multicolumn{4}{c}{$r_{1,6}$} &
		\multicolumn{6}{c}{--} \\
		\multicolumn{12}{c}{$Q_2 \to Q_3 \to Q_4 \to Q_5 \to Q_6 \to Q_7$} &
		\multicolumn{4}{c}{1.04} &
		\multicolumn{4}{c}{$r_{2,7}$} &
		\multicolumn{6}{c}{-0.70 (0.26)} \\
		\midrule
		\multicolumn{12}{c}{$Q_1 \to Q_2 \to Q_3 \to Q_4 \to Q_5 \to Q_6 \to Q_7$} &
		\multicolumn{4}{c}{1.24} &
		\multicolumn{4}{c}{$r_{1,7}$} &
		\multicolumn{6}{c}{--} \\
		\bottomrule
		\bottomrule
	\end{tabular*}
\end{table*}

\begin{table*}[cols=28,pos=h,width=.8\textwidth]
	\caption{The detailed information of three-qubit Motion-CPMG experiment.}\label{table_3}
	\begin{tabular*}{\tblwidth}{@{} LLLLLLLLLLLLLLLLLLLLLLLLLLLL@{} }
		\toprule
		\toprule
		\multicolumn{12}{c}{Qubit ID} &
		\multicolumn{6}{c}{$Q_2$} &
		\multicolumn{4}{c}{$Q_3$} &
		\multicolumn{6}{c}{$Q_4$} \\
		\midrule
		\multicolumn{12}{c}{$T_1~(\mu s)$} &
		\multicolumn{6}{c}{12.8} &
		\multicolumn{4}{c}{11.6} &
		\multicolumn{6}{c}{10.3} \\
		\multicolumn{12}{c}{$T_2^*~(\mu s)$} &
		\multicolumn{6}{c}{0.75} &
		\multicolumn{4}{c}{0.87} &
		\multicolumn{6}{c}{0.23} \\
		\multicolumn{12}{c}{$\tau_{CPMG-2}~(\mu s)$} &
		\multicolumn{6}{c}{3.46} &
		\multicolumn{4}{c}{3.61} &
		\multicolumn{6}{c}{--} \\
		\multicolumn{12}{c}{$\tau_{CPMG-4}~(\mu s)$} &
		\multicolumn{6}{c}{--} &
		\multicolumn{4}{c}{--} &
		\multicolumn{6}{c}{3.68} \\
		\midrule
		\midrule
		\multicolumn{12}{c}{Motion-CPMG} &
		\multicolumn{6}{c}{$\tau_L~(\mu s)$} &
		\multicolumn{4}{c}{correlation} &
		\multicolumn{6}{c}{$r_{i,j} (\pm)$} \\
		\midrule
		\multicolumn{12}{c}{$Q_2~CPMG-1 \to Q_3~CPMG-1$} &
		\multicolumn{6}{c}{4.49} &
		\multicolumn{4}{c}{$r_{2,3}$} &
		\multicolumn{6}{c}{0.24 (0.02)} \\
		\multicolumn{12}{c}{$Q_3~CPMG-1 \to Q_4~CPMG-2$} &
		\multicolumn{6}{c}{5.34} &
		\multicolumn{4}{c}{$r_{3,4}$} &
		\multicolumn{6}{c}{-0.07 (0.02)} \\
		\multicolumn{12}{c}{$Q_2~CPMG-1 \to Q_3~CPMG-1 \to Q_4~CPMG-2$} &
		\multicolumn{6}{c}{5.87} &
		\multicolumn{4}{c}{$r_{2,4}$} &
		\multicolumn{6}{c}{-0.01 (0.05)} \\
		\bottomrule
		\bottomrule
	\end{tabular*}
\end{table*}

\end{document}